\documentclass[useAMS,usenatbib]{mnras}

\usepackage{graphicx}
\usepackage{amssymb}
\usepackage{amsmath}
\usepackage{datetime}
\usepackage[normalem]{ulem}
\usepackage{times}
\usepackage{bm} 
\usepackage{comment}

\usepackage{color}
\usepackage{soul}
\definecolor{stan}{rgb}{0,0,1}

\definecolor{xyz}{rgb}{0.431,0.106,0.537}

\definecolor{asif}{rgb}{1.0,0.1,0.0}

\date{\currenttime \today}

\begin{document}

\title[Oblique Rotators]
{
3D MHD models of the centrifugal magnetosphere from a massive star with an oblique
dipole field
}

 \author[A. ud-Doula et al.]
{
Asif ud-Doula$^1$\thanks{Email: asif@psu.edu},
Stanley P. Owocki$^2$, Christopher Russell$^2$,
Marc Gagn\'e$^3$ and
\newauthor
Simon Daley-Yates$^4$
\\
 $^1$ Penn State Scranton, Dunmore, PA 18512, USA.\\
 $^2$ Bartol Research Institute, Department of Physics and Astronomy, 
 University of Delaware, Newark, DE 19716, USA\\
 $^3$Department of Earth and Space Sciences, West Chester University, PA, USA\\
 $^4$ School of Physics and Astronomy, University of St Andrews, North Haugh, St Andrews, Fife, Scotland KY16 YSS, UK
}

\def\<<{{\ll}}
\def\>>{{\gg}}
\def\wig{{\sim}}
\def\spose#1{\hbox to 0pt{#1\hss}}
\def\ltwig{\mathrel{\spose{\lower 3pt\hbox{$\mathchar"218$}}
     R_{\rm A}ise 2.0pt\hbox{$\mathchar"13C$}}}
\def\gtwig{\mathrel{\spose{\lower 3pt\hbox{$\mathchar"218$}}
     R_{\rm A}ise 2.0pt\hbox{$\mathchar"13E$}}}
\def\+/-{{\pm}}
\def\=={{\equiv}}
\def\mubar{{\bar \mu}}
\def\mustar{\mu_{\ast}}
\def\Lambar{{\bar \Lambda}}
\def\Rstar{R_{\ast}}
\def\Mstar{M_{\ast}}
\def\Lstar{L_{\ast}}
\def\Tstar{T_{\ast}}
\def\gstar{g_{\ast}}
\def\vth{v_{th}}
\def\grad{g_{rad}}
\def\glines{g_{lines}}
\def\Mdot{\dot M}
\def\mdot{\dot m}
\def\yr{{\rm yr}}
\def\ksec{{\rm ksec}}
\def\kms{{\rm km/s}}
\def\qad{\dot q_{ad}}
\def\qlines{\dot q_{lines}}
\def\solar{\odot}
\def\Msun{M_{\solar}}
\def\msbyr{\Msun/\yr}
\def\Rsun{R_{\solar}}
\def\Lsun{L_{\solar}}
\def\Be{{\rm Be}}
\def\Rpole{R_{p}}
\def\Req{R_{eq}}
\def\Rmin{R_{min}}
\def\Rmax{R_{max}}
\def\Rstag{R_{stag}}
\def\vinf{V_\infty}
\def\Vrot{V_{rot}}
\def\Vcrit{V_{\rm crit}}
\def\half{{1 \over 2}}
\newcommand{\beq}{\begin{equation}}
\newcommand{\eeq}{\end{equation}}
\newcommand{\beqa}{\begin{eqnarray}}
\newcommand{\eeqa}{\end{eqnarray}}
\def\phip{{\phi'}}

\maketitle

\begin{abstract}
We present results from new  self-consistent 3D MHD simulations of the magnetospheres from massive stars with a dipole magnetic axis that has a non-zero obliquity angle ($\beta$) to the star's rotation axis.  As an initial direct application, we compare the global structure of co-rotating disks for nearly aligned ($\beta=5^o$) versus half-oblique ($\beta=45^o$) models, both with
moderately rapid  rotation ($\sim$ 0.5 critical). 
We find that accumulation surfaces broadly resemble the forms predicted by the analytic Rigidly Rotating Magnetosphere (RRM) model, 
but the mass buildup to near the critical level for centrifugal breakout against magnetic confinement distorts the field from the imposed initial dipole. This leads to an associated warping of the accumulation surface toward the rotational equator, with the highest density concentrated in {\em wings} centered on the intersection between the magnetic and rotational equators.
These MHD models can be used to synthesize rotational modulation of photometric absorption and H$\alpha$ emission for a direct comparison with observations.
\end{abstract}

\begin{keywords}
(magnetohydrodynamics) MHD ---
Stars: winds, outflows ---
Stars: magnetic fields ---
\end{keywords}

\section{Introduction}

Hot luminous, massive stars of spectral type O and B have dense, high-speed, radiatively driven stellar winds
\citep{Cas1975}.
In the subset ($\sim$10\%; \cite{2017MNRAS.465.2432G,2019MNRAS.483.2300S}) of massive stars with strong ($>\,100$\,G; \cite{2019MNRAS.482.3950S}), globally ordered (often significantly dipolar; \citet{2019A&A...621A..47K}) magnetic fields, the trapping of this wind outflow by closed magnetic loops leads to the formation of a circumstellar {\em magnetosphere} \citep{2013MNRAS.429..398P}.
Because of the large angular momentum loss associated with 
their relatively strong, magnetised wind 
\citep{udD2009}, magnetic O-type stars are typically
slow rotators,  with trapped wind material falling back on a dynamical timescale, giving what's known as a ``{\em dynamical magnetosphere}" (DM).

However, in the case of magnetic B-type stars, such angular momentum loss is greatly  reduced due to their relatively weak stellar winds, implying longer spin-down times. Thus, not surprisingly, a significant fraction of B-type stars still retain a moderately rapid rotation. For such cases, the associated Keplerian co-rotation radius $R_{\rm K}$ lies within the Alfv\'{e}n radius $R_{\rm A}$ that characterises the maximum extent of the magnetosphere.  The rotational 
support within the magnetosphere leads to formation of a ``{\em centrifugal magnetosphere}'' (CM), 
wherein the much longer confinement time allows material to build up to much higher density than in DM's.

For the special case of a dipole field aligned with the rotation axis, \citet{udD2008} %ud-Doula, Owocki and Townsend (2008) 
carried out 2D MHD simulations of the resulting axisymmetric CM with material concentrated along the common rotational and magnetic equator.
The central aim of the current paper is to present results from new 3D MHD simulations for the sample case of an oblique dipole field that has a tilt angle $\beta=45^o$, and characterize its more complex, inherently 3D CM.

For models without rotation, initial 2D MHD simulations of such wind-fed magnetospheres by \citet{udDOwo2002} %ud-Doula and Owocki (2002)
showed that, for a star with radius $R_\ast$ and dipole field of surface strength $B_{\rm eq}$
at the magnetic equator, the competition of the field with a stellar wind of mass loss rate ${\dot M}$ and terminal speed $V_\infty$ can be characterized in terms of a dimensionless, wind-magnetic-confinement parameter $\eta_\ast \equiv B_{\rm eq}^2 R_\ast^2/{\dot M} V_\infty$, with the Alfv\'{e}n radius then scaling as $R_{\rm A} \approx R_\ast \eta_\ast^{1/4}$.
The follow-on study of aligned rotation by \citet{udD2008}
%ud-Doula et al. (2008)
parameterized its effects by the dimensionless ratio $W \equiv V_{\rm rot}/V_{\rm orb}$ between the equatorial rotation speed and the orbital speed near the equatorial surface.
In the inner region where the field maintains rigid-body rotation, this gives a Kepler co-rotation radius $R_{\rm K} = R_\ast W^{-2/3}$, at which gravitational and centrifugal forces are in balance.

The observational compilation by \citet[][see in particular their figure 3]{2013MNRAS.429..398P} %Petit et al. (2013; see in particular their figure 3) 
shows that many B-stars have $R_{\rm A} > R_{\rm K}$, with confinement parameters ranging even to $\eta_\ast > 10^6$.
For example, the prototypical CM star $\sigma$~Ori~E, has $\eta_\ast \approx 10^6$ and $W \approx 0.34$, implying $R_{\rm A}/R_\ast \approx 31 \gg R_{\rm K}/R_\ast \approx 2.1$,
and thus an extensive CM.

The field stiffness and associated high Alfv\'{e}n speed of a star with such large $\eta_\ast$ imply very small Courant time step
in direct MHD simulations, and so far this has limited MHD models to $\eta_\ast \lesssim 10^3$.
Alternatively, by considering the limit of an arbitrarily strong field (effectively with $R_A \rightarrow \infty$), semi-analytic analyses based on an idealization of purely rigid fields have led to 
a {\em rigidly rotating magnetosphere} (RRM) formalism \citep{Tow2005},
%(Townsend and Owocki 2005), 
which derives how CM material accumulates on surfaces set by minima of a combined centrifugal and gravitational potential, under the assumed condition of rigid-body rotation.
Such RRM models have shown great potential for explaining key observational signatures of CM stars, e.g. rotational modulation of Balmer line emission
\citep{Tow2005, 2012MNRAS.419..959O}.
%(Townsend, Owocki and Groote 2006; Oksala 2012).

However, central weakness of this RRM approach is that it provides no description for how the stellar-wind-fed accumulation surfaces of the CM are ultimately emptied.
Recent theoretical analyses \citep{2020MNRAS.499.5366O,2022MNRAS.513.1449O},
%(Owocki et al. 2020, 2022), 
developed to explain empirical scaling for H-alpha
\citep{2020MNRAS.499.5379S}
%(Shultz et al. 2020) 
and radio emission \citep{2021MNRAS.507.1979L,2022MNRAS.513.1429S}
%(Leto et al. 2021; Shultz et al. 2022) 
from CM stars, provide strong evidence that this emptying occurs through frequent, low-level, {\em centrifugal breakout} (CBO) events.
These are triggered when the accumulated mass exceeds a critical level for which the centrifugal force overwhelms the confinement of the magnetic field tension force.
Applying the estimated CBO critical density distribution within the RRM formalism for accumulation surfaces, \citet{2022MNRAS.511.4815B}
%Berry et al. (2022) 
recently modeled the photometric light variation associated with absorption and scattering emission from magnetic clouds around the prototypical CM star $\sigma$~Ori~E.

The CBO-limited density distribution derived by
\citet{2020MNRAS.499.5366O}
%Owocki et al. (2020)
was actually based on analysis for the simplified special case of an aligned dipole, calibrated against the 2D MHD simulations by \citet{udD2008}.
%ud-Doula et al. (2008).
Specifically, assuming $W=1/4$ or 1/2 and the strongest allowed confinement $\eta_\ast=1000$, the MHD models show that, in the CM region extending above $R_{\rm K}$, the critical surface density accumulated along the common magnetic and rotational equators declines in radius as $r^{-6}$, consistent with the analytic CBO analysis that this should follow the decline in magnetic tension $\sim B^2$.

But for the many CM's with a nonzero tilt angle $\beta$ between the field and rotation axes,
it is not clear how the density on the accumulation surface should vary in {\em azimuth} away from the direction set by intersection of the rotational and magnetic equators.

The paper here presents new 3D MHD simulations of the CM formed by a tilted dipole with axis that makes an angle $\beta=45^o$ with the rotation axis.
The standard parameterization of moderately rapid rotation ($W=1/2$) and strong confinement
($\eta_\ast = 10^3$) provides, for the first 
time, a 3D MHD, oblique-dipole model with an extended CM region\footnote{By comparison, the recent 3D simulations by 
\citep{2022MNRAS.515..237S}
are limited to modest $\eta_\ast = 50$ with $R_{\rm A} \approx 2.7 R_\ast \gtrsim R_{\rm K}$ and have too small CM regions to enable direct comparison with RRM models.}, 
ranging here from $R_K \approx 1.6 R_\ast$ to $R_{\rm A} \approx 5.6 R_\ast$.
A particular emphasis is to characterize the resulting 3D, dynamical distribution of density, and compare that with expected
accumulation surfaces from the semi-analytic RRM model.
A specific goal is to test and calibrate the RRM density parameterization used by \citet{2022MNRAS.511.4815B},
%Berry et al. 2022, 
which was inspired by the CBO analysis and preliminary versions of the 3D MHD simulations presented here.

To lay the basis for results presented in Section 3, Section 2 first reviews the general numerical MHD approach, numerical grid, boundary conditions and  parameter domain. Section 4 concludes with a summary and outline for future work.

\vskip 0.2in

\section{Numerical Setup and Initial Conditions}

\begin{comment}
\begin {itemize}
\item Massively parallel PLUTO code
\item $\zeta$~Pup stellar parameters
\item $\eta_\ast = 1000$ , $\beta=5^o, 45^o , 90^o$, $w=0.5 w_{critical}$; 
$R_{Alf} = 5.5R_*$; $R_{Kep} =1.6 R_*$
\item  3D spherical grid: nr=250; nq=64; nf=128
\item radial grid logarithmic, uniform in co-latitude and azimuth
\item isothermal equation of state
\item Inflow/outflow boundary condition at the lower boundary
\item ÔRing averageÕ around Z-axis to avoid coordinate singularity
\item Hyperbolic Divergence Cleaning (eGLM) to enforce divergence free constraint
\item HLL solver with Linear reconstruction and RK2 time integrator
\item Background magnetic field splitting

\end{itemize}
\end{comment}
Most of our previous numerical models were performed using {\sc Zeus-3D} or {\sc Zeus-MP} codes, but here we use publicly available, massively parallel MHD code {\sc Pluto} (version 4.4) \citep{Mignone2007},
because of its highly versatile, modular structure 
that is well suited for modern Linux clusters.

The winds of massive stars are highly ionized and the competition between photoionization heating and radiative cooling keeps the wind close to the stellar effective temperature \citep{Pau1987,Dre1989}. As such, we can approximate the wind to be isothermal and model it with standard magnetohydrodynamics (MHD) equations in {\rm cgs} units:
\begin{equation}
\label{eq:mass}
\frac{\partial \rho}{\partial t} + \bm{\nabla} \cdot \left( \rho \bm{v} \right) = 0
\end{equation}

\begin{equation}
\label{eq:momentum}
\frac{\partial {v}}{\partial t} + \left( \bm{v} \cdot  {\nabla} \right) \bm{v}
+ \frac{1}{4 \pi \rho} \bm{B} \times  \left( \nabla \times \bm{B} \right)
+ \frac{1}{\rho} \nabla p = \bm{g} + \bm{g}_{\mathrm{lines}} + \bm{F}_{\mathrm{co}}
\end{equation}
\begin{equation}
\label{eq:magnetic}
\frac{\partial \bm{B}}{\partial t}
+ \nabla \times \left( \bm{B} \times \bm{v} \right) = 0.
\end{equation}
Here
$\rho$, $\bm{v}$, $\bm{B}$, $p$, $\bm{g}$ and
$\bm{g}_{\mathrm{lines}}$
are, the density, velocity, magnetic field, pressure,
and accelerations due to respectively gravity and line-scattering of radiation. 
The comoving frame acceleration 
$\bm{F}_{\mathrm{co}}$ is the sum of both the centrifugal and Coriolis forces, given respectively by 
\begin{equation}
\bm{F}_{\mathrm{centrifugal}} =
- \left[ \bm{\Omega}_{\mathrm{fr}} \times \left( \bm{\Omega}_{\mathrm{fr}}
\times \bm{R} \right) \right]
\end{equation}
and
\begin{equation}
\bm{F}_{\mathrm{coriolis}} =
- 2 \left( \bm{\Omega}_{\mathrm{fr}} \times \bm{v} \right) \, ,
\end{equation}
where $\bm{\Omega}_{\mathrm{fr}}$ is the angular frequency of the rotating frame with $\bm{r}$ the radial distance vector.

As the wind is assumed to be isothermal 
at the stellar surface temperature $T$, 
we close equations (\ref{eq:mass} - \ref{eq:magnetic}) using 
the ideal gas equation of state,
\begin{equation}
p = \frac{\rho k_B T}{\mu} 
= \rho c_{\mathrm{iso}}^{2},
\end{equation}
where $k_{\mathrm{B}}$ is the Boltzmann constant,
$\mu=0.6 m_p$ is the molecular weight, and the last equality casts this in terms of the isothermal sound speed $c_{\mathrm{iso}}$.

\subsection{Radial line-driving of wind outflow}

The radial outflow described in the previous section arises from the strong radial driving of the line-force,
$\bf{g}_{\rm{lines}}$.
As in \citet{udDOwo2002}, we model this here in terms of the
standard  \citet*[hereafter CAK]{Cas1975} formalism,
corrected for the finite cone angle of the star, using
a spherical expansion approximation for the local flow gradients
\citep*{Pau1985,FriAbb1986}
and ignoring {\em non-radial} line-force components that can
arise in a non-spherical outflow.
Although such
non-radial terms are typically only a few percent of the
radial force \citep{OwoudD2004}, in non-magnetic models of rotating winds, they act without much
competition in the lateral force balance, and so can have surprisingly
strong effects on the wind channeling and rotation
\citep*{OwoCra1996,GayOwo2000}.
But in magnetic models with an already strong component of non-radial
force, such terms are not very significant, and
since their full inclusion substantially complicates both the
numerical computation and the analysis of simulation results, we have elected to
defer further consideration of such non-radial line-force terms to future studies.

By limiting our study to moderately fast rotation, half
or less of the critical rate, we are also able to neglect the effects
of stellar oblateness and gravity darkening.

\subsection{Simulation}
\label{sec:sim}
For our numerical scheme, we choose 
a method that is
fully unsplit and 2nd order accurate in space and time, 
using linear reconstruction, Runge-Kutta time stepping and  
the HLL Riemann solver. The extended GLM divergence cleaning algorithm was used to ensure the $\nabla \cdot \bm{B} = 0$ condition. For highly magnetized plasma, such as the ones discussed here, it is advantageous to use a background magnetic field (typically dipolar, as is the case here) and  evolve its deviation in time rather than the actual magnetic field. 

\subsection{Numerical grid}
\label{sec:grid}
For all our models, we use a stretched rectilinear spherical polar grid extending from $r=R_\ast$ to $r=25 R_\ast$ in which the physical volume is discretised with 250 cells in $r$, 64 cells in $\theta$ and 128 cells in $\phi$. This leads to a cell size in the $r$ direction which stretches from $\Delta r_{1}~\approx~0.008 \ R_{\ast}$ to $\Delta r_{250}~\approx~0.542\ R_{\ast}$ with a constant stretching factor of $1.023$. Both the $\theta$ and $\phi$ directions have uniform spacing. The stretching regime in the radial direction is required to resolve the sonic point of the wind, which is very close to the stellar surface. Typically, at least 5-10 grid points are required to resolve the sonic point to ensure accurate base mass flow. 

\subsection{Boundary conditions}
\label{sec:BCs}
Boundary conditions are  challenging in MHD modelling and great care must be taken to avoid any unphysical outcomes. 
For the most part, we closely follow boundary conditions outlined in \citet{udDOwo2002} and \citet{Daley-Yates2019}, with the latter describing the first obliquely rotating massive-star winds, although their focus was on radio emission from such objects. Similar boundary conditions are also employed by
\citet{2022MNRAS.515..237S}
albeit for the geodesic mesh-based {\sc Riemann Geomesh} code.

In brief, the outer radial boundary of all our simulations is set to outflow, which allows material to freely leave the computational domain. The inner radial boundary is set to `inflow' such that  the star is continually feeding material to the wind and therefore replenishing material in the simulation. 

The velocity in the lower radial boundary is specified by linearly extrapolating back from the first 2 computational cells above the boundary, allowing the flow into the computational active zone to adjust to the conditions of the wind and permitting material to also re-enter the stellar surface as magnetically confined material follows field lines back to the stellar surface. We limit the maximum radial inflow/outflow speeds to the 
fixed sound speed. 
Specifying the boundary in this manner also allows the mass loading of the wind to self consistently adapt to the rotation of the star. Large rotational velocities can impact the mass-loss of a star. This is due to the effective gravity at the rotational equator being reduced relative to the poles, leading to material being lifted from the surface more easily.

The boundary of the lower and upper azimuthal direction is assumed to be periodic. The upper and lower boundary of the polar direction was set to reflective so as not to act as a sink for material. This final boundary condition is non-physical and a reflective polar boundary can lead to spurious heating or jets along the polar axis. There are several methods designed to overcome this numerical difficulty. One such method is known as $\pi$-boundary conditions in which the fluid quantities are translated $\pi$ around the axis and vector values transformed such that material effectively passes over the pole. This method is implemented in the public codes {\sc Athena++} \citep{White2016} and {\sc MPI-AMRVAC} \citep{Xia2018}. {\sc Pluto} provides a similar functionality called `polaraxis' which we utilize in our simulations here.

Since we use an isothermal equation of state for the wind, we neglect behaviour due to both shock heating and radiative cooling,
both of which can
play a role in the wind dynamics \citep{udD2008, udD2013}. 
Inclusion of a full energy balance with radiative cooling is thus a goal for future studies.

\subsection{Stellar Parameters}

We performed a number of simulations both in 3D and 2D to ensure the results are consistent. Here, we focus specifically on 3D models with inclination angle between the magnetic field and rotation axes of 45$^o$ and 5$^o$;
the latter mimics a 
field-aligned model, but with a small inclination
to ensure azimuthal symmetry is numerically broken.

Following previous studies \citep{udDOwo2002,udD2008}, we use the stellar parameters of $\zeta$ Pup, a prototypical O-supergiant. In the absence of magnetic field, its mass loss is assumed to be about $3.0 \times 10^{-6} M_\odot /$yr. For our model here we assume a dipolar magnetic field with polar strength of 9300 G corresponding to magnetic confinement of $\eta_\ast =1000$. 
Although these differ from the  parameters of a typical Bp star, %nonetheless 
the models still mimic general trends in magnetospheric structure with the key magnetic confinement parameter \citep{udDOwo2002}.

\begin{figure}
\includegraphics[width=0.58\textwidth]{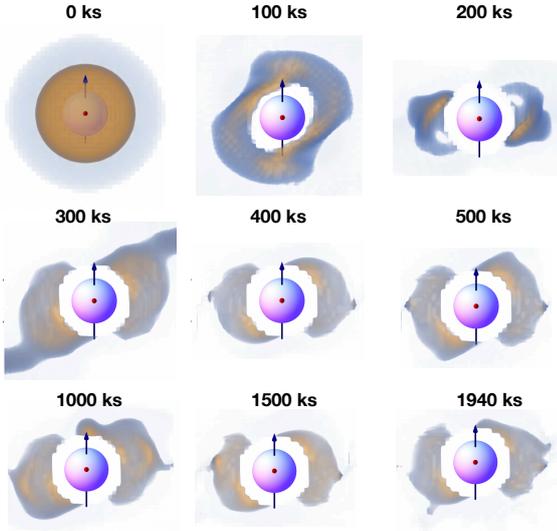}
   \caption{Time evolution of the standard model ($\eta_\ast=10^3$, $W=1/2$, $\beta=45^o$),
   showing volume renderings of the density structure as viewed from the rotational pole (red dot), with projected magnetic axis (blue arrow) directed upward.
   After initial transits, the dragonfly wing-like co-rotating structure settles into a quasi-steady state with occasional outbreaks due to centrifugal forces.
.}
    \label{fig:1}
\end{figure}

Our assumed rotation is half the critical, corresponding to about 250 km/s, with axis aligned with z-axis. 
Following the approach of \citet{Daley-Yates2019},
the magnetic pole is rotated about $y$-axis, and so lies in $xz$-plane, as indicated by the blue arrow in figures \ref{fig:1} - \ref{fig:Bfield}.

\subsection{Initial conditions}
\label{sec:IC}

The initial conditions of the simulations are specified using the density and velocity profile equations assuming a spherically symmetric wind with CAK mass loss rate \citep{OwoudD2004}  and `$\beta$'-law, i.e.  $v(r)=v_\infty (1-R_\ast/r)^\beta$ with $\beta=0.8$ and $v_\infty \propto v_{\rm escape}$.

The magnetic field is initialised as an ideal dipole, centred at the origin and rotated about the $y$-axis, in the $xz$-plane. We then evolve the model by letting the wind and magnetic field compete against each other. 

\section{Results}
\subsection{Time Evolution of the Standard Model}

To characterize our 3D models we use a combination of 2D slices and 3D projection plots.
For our standard model with dipole tilt angle $\beta=45^{\rm o}$, figure \ref{fig:1} shows for example the time evolution of 3D density as viewed from an inclination $i=0$ over the rotational pole (here marked  by a red dot, indicating a rotational vector pointing directly toward the observer).
Since we are primarily interested in the 3D structure of the CM that forms above $R_{\rm K}$, we have for illustrative clarity chosen 
to hollow out the density for radii $r < R_{\rm K}$,  ignoring the DM part of the magnetosphere which has been extensively discussed in previous studies  (e.g. see, \citet{udD2008,udD2013}).

The blue arrow pointing upward marks the magnetic dipole axis, with the projection of the magnetic equator thus along the horizontal.
Starting from the initial condition of a spherical outflow, the combination of rotation and strong field confinement progressively channels material toward greater concentration, forming two opposing {\em wings} that straddle the common rotational and magnetic equator.
This wing structure is already clear in the 400\,ks snapshot, after which there are relatively modest variations about this basic shape, extending here to nearly 5 times longer, to the well-relaxed, final simulation time of 1940\,ks.
By comparison, for a typical wind speed of $V_w \approx$\,1000\, km/s, the dynamical wind crossing time through a stellar radius $R_\ast$ is just $t_w = R_\ast/V_w \approx 14$\,ks.

\subsection{Contrast with nearly aligned case: Disk vs.\ Wings}

\begin{figure}
%    \centering
\includegraphics[width=0.68\textwidth]{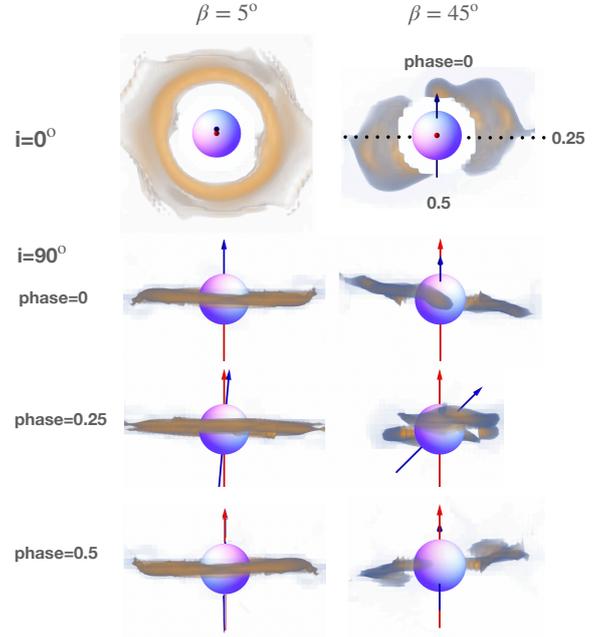}
\caption{As in figure \ref{fig:1}, volume renderings of density structure, now comparing representative evolved time step ($t=1000$\,ks) structures of two different models with identical magnetic field, rotation but different  obliquity, $\beta$ : 45$^o$ (right) and 5$^o$ (left). Notice how higher obliquity model restricts density in the azimuthal direction. This is a result of complex dynamics between rotation and strong tilted magnetic field.}
    \label{fig:2s}
\end{figure}

To highlight further the distinct wing structure formed for this $\beta=45^{\rm o}$ oblique dipole, figure \ref{fig:2s} compares a representative evolved time (right column) with a similarly evolved model for the nearly aligned case $\beta=5^{\rm o}$ (left column).
The top row again shows the view from inclination $i=0$ over the rotational pole, with the dipole axis upward and the dotted horizontal lines marking the intersection between the magnetic and rotation equators.

The nearly aligned model on the left has its density in a nearly azimuthally symmetric, equatorial disk.
In contrast, the oblique dipole on the right has density that is azimuthally concentrated in wings that straddle 
the common magnetic/rotation equator.
Note, however, the modest {\em prograde} asymmetry, with a somewhat higher density in the direction {\em toward} the stellar rotation (here counter-clockwise).

The bottom 3 rows show equatorial views ($i=90^{\rm o}$) at 3 rotational phases,  corresponding to the dipole axis inclination toward (phase=0), perpendicular to (phase=0.25),
and away from (phase=0.5) the observer.
These show that the wings are quite near the rotational equator, but with a distinct warping out of the equatorial plane at larger radii.
For phase=0.25, showing views along the common equator, the
foreground wing partially obscures the stellar disk, but with a mark upward offset from this warping.
At phase=0, the prograde extension of high density also leads to some obscuration;
but a half-period later, 
at phase=0.75, there is now little material in front of the star, since the view now through the trailing gap between the two wings.

Clearly, such details will have important consequences for rotationally modulated light curves from stars with tilted CM's.

\begin{figure}
    \centering
\includegraphics[width=0.48\textwidth]{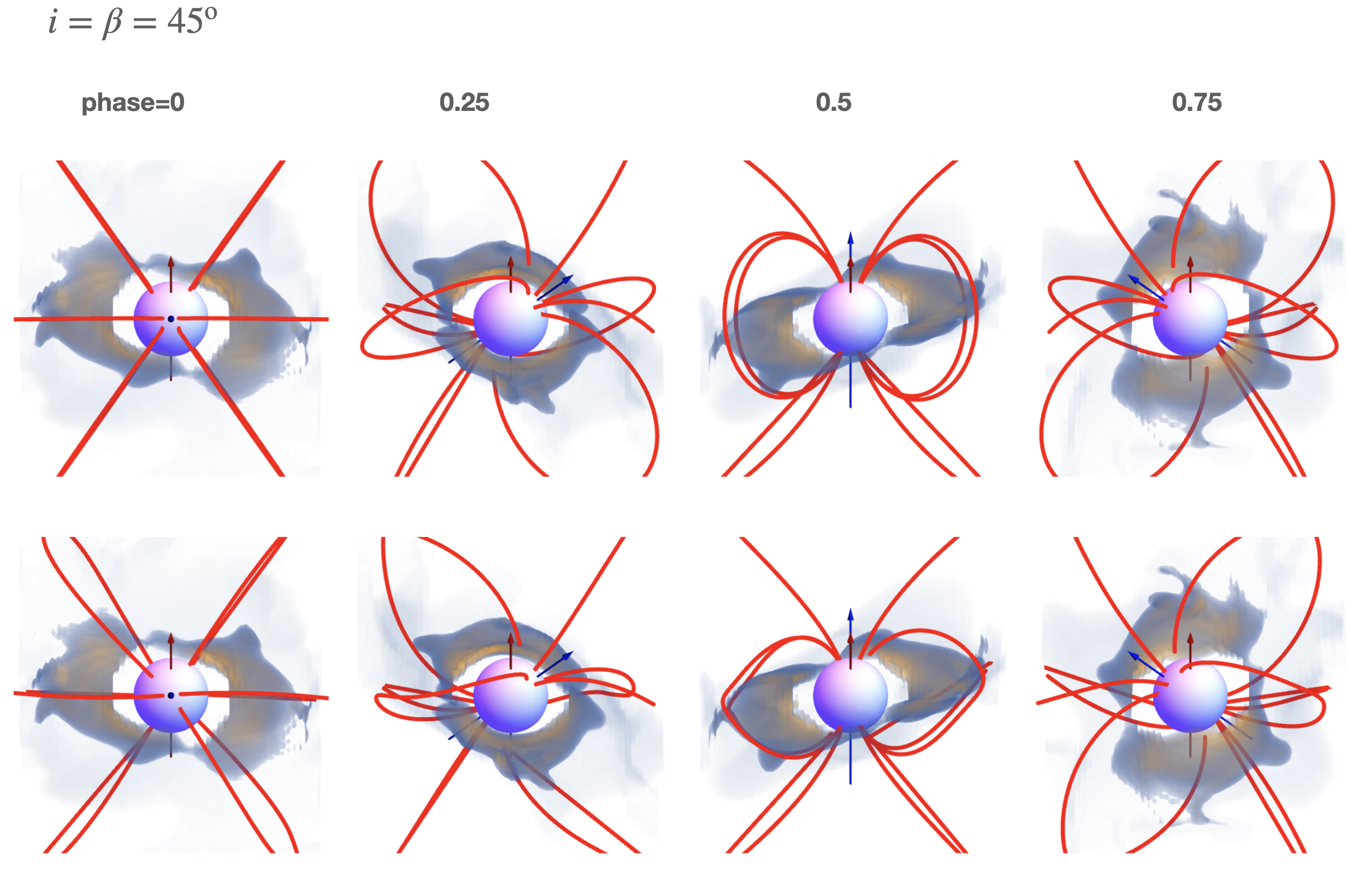}
   \caption{For an observer with inclination $i=45^o$, the red curves compare the magnetic field lines for the initially imposed dipole  (top) versus the dynamically evolved field at the final time step (bottom).
   These are both superposed on
   the associated phase variation of volume-rendered density for this final time step to highlight the distortion of the magnetic field due to wind dynamics.
}
    \label{fig:Bfield}
\end{figure}

\begin{figure}
\includegraphics[width=0.23\textwidth]{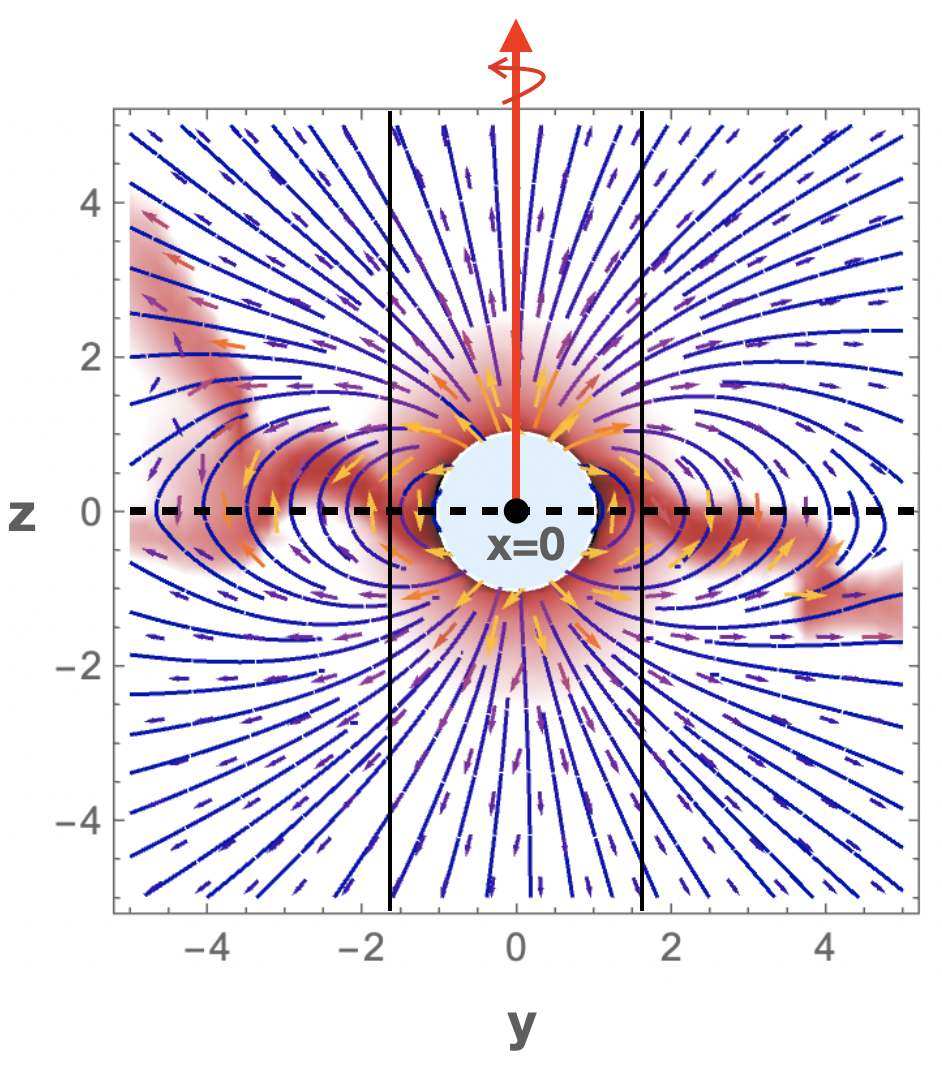}
\includegraphics[width=0.25\textwidth]{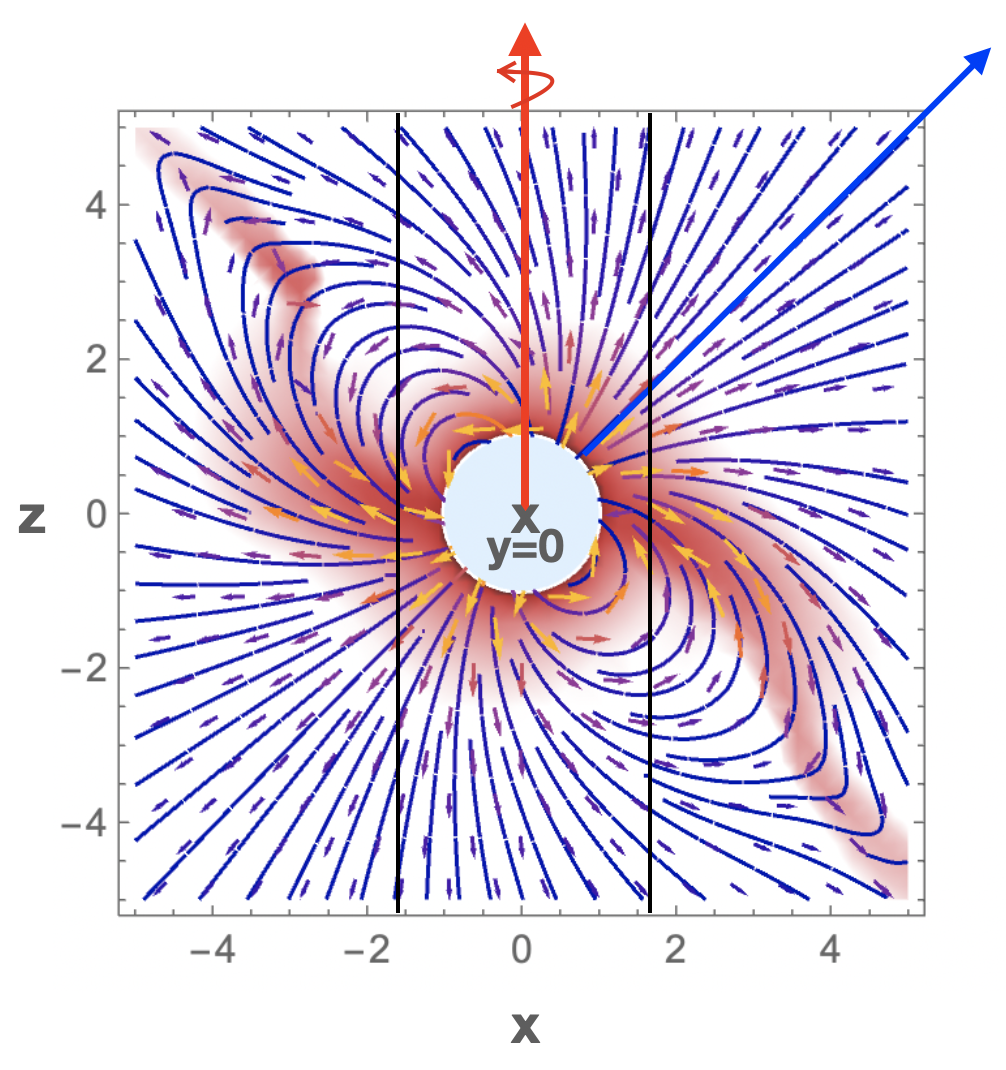}
\caption{For the final time snapshot $t=1940$\,ks of the $\beta=45^o$ tilted dipole, 2D slices showing density, field lines, and mass flux vectors in the $x=0$ (left) and $y=0$ (right) planes, where $y$ is along the common equator (dashed line), $z$ is the rotation axis, and the field tilt is in the $xz$ plane. The blue vector in the right panel shows the dipole axis in this $xz$ plane.
The vertical lines mark offsets from the rotation axis by one Kepler radius $R_{\rm K}$.}
    \label{fig:y0x0}
\end{figure}

\begin{figure*}
\centering
\includegraphics[width=1.0\textwidth]{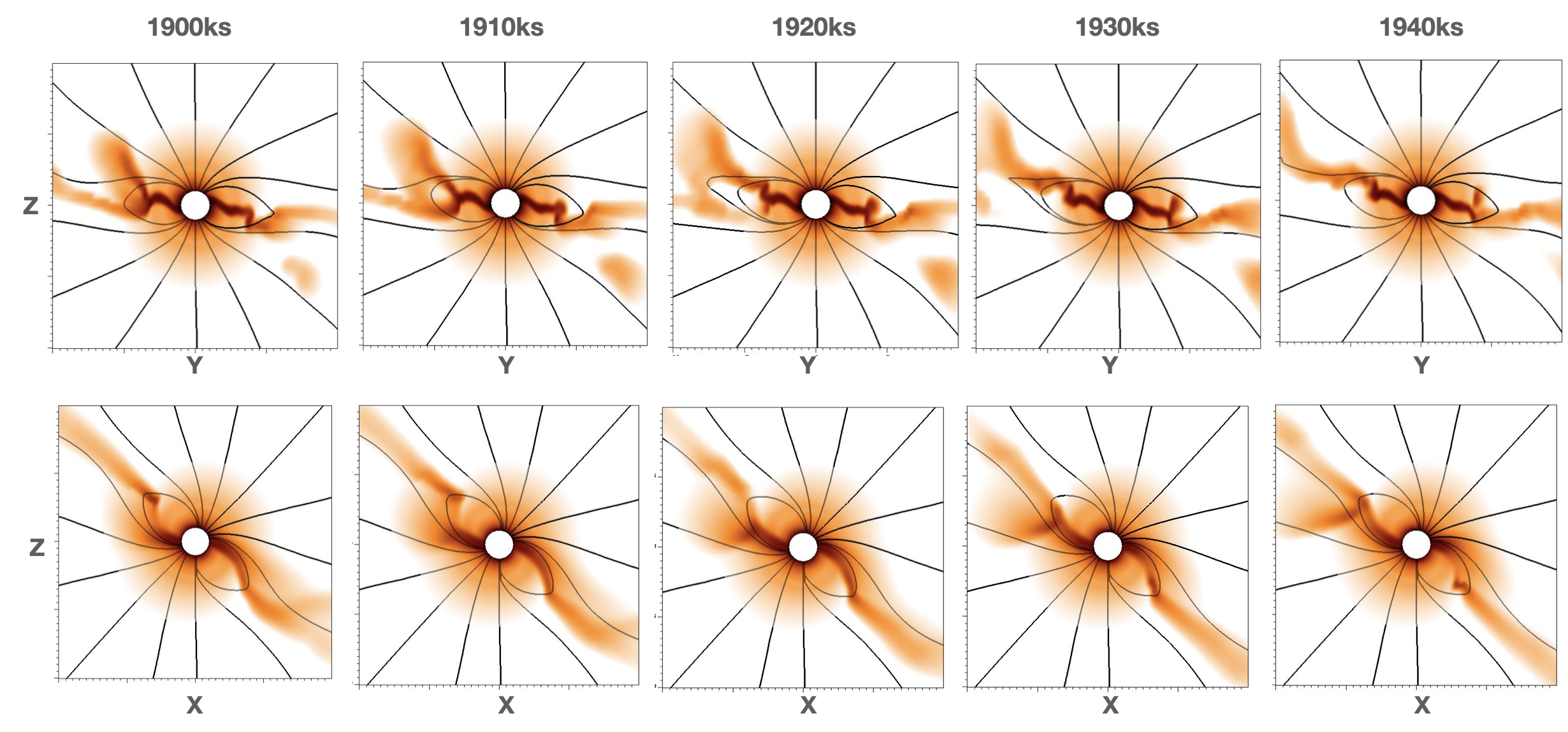}
   \caption{
   Time evolution over the last five snapshots ($t=1900-1940$\,ks) of log density and field lines, showing both yz (top) and xz (bottom) slices, now extending over spatial range of $\pm 10 R_\ast$.
   In the yz cut note the progressive field line stretching and mass ejection associated with a centrifugal breakout event. Breakouts also occur in the xz cut but they are more conspicuous in a different snapshot due to slight left-right asymmetry in the numerical model. Evidence of such breakout events are quite apparent in figure \ref{fig:7} as well and occur on about 50 ks timescale
.}
    \label{fig:ZXZYlast50}
\end{figure*}

\subsection{Centrifugal distortion of dipole field by CM plasma}

This warped-wing structure results from trapping of centrifugal material by the strong magnetic field, which in turn distorts the field from its dipole form.
To illustrate this, figure \ref{fig:Bfield} adds magnetic field lines (red curves) to the density surfaces, as viewed now from observer inclination $i=45^{\rm o}$ at the labeled rotational phases.
The upper panel shows the initial tilted dipole, while the lower panel shows the MHD dynamical field at the final time.
While the two are quite similar, the differences show the result of the distortions of the field from the CM plasma, 
in particular the centrifugal stretching from the trapped material in the dense wings.

For example, at phase=0, which gives a view looking down the magnetic pole, the dipole field lines project into simple X-cross, whereas the dynamical fields show a rotational twist between lines into and out of the page.

At phase=0.5, note that closed dipole loops in the upper panel show an outward stretching in the lower panel. Such stretching of the magnetic field leads to eventual centrifugal breakout of the trapped material. 

Comparison of phase=0.25 and 0.75 show a simple left/right swap for the dipole fields;
but there is an asymmetric distortion in the closed dynamical field lines, reflecting a notable asymmetry in the density structure as well.

\subsection{2D slices in zy and zx planes}

To complement this 3D rendition of the field, which can be difficult to track visually, figures \ref{fig:y0x0} and \ref{fig:ZXZYlast50} show 2D slices of the density and field line structure in the $yz$ ($x=0$) and $xz$ ($y=0$) planes.
The mass flux vectors ($\rho {\bf v}$) in figure \ref{fig:y0x0}
show how magnetically channeled outflow compresses material into high density structures.
The vertical, z-axis is along the rotation vector (red arrows), while the y-axis is along the common magneto-rotational equator, marked by the horizontal dashed line in the left panel for $x=0$.
The closed loops in this plane confine the dense plasma against centrifugal forces, forming the center of the dense wing structure in 3D.
By contrast, in the right panel with $y=0$, mass flux along equatorward boundary of the closed to open field channels material along a diagonal to the loop tops. As shown in the outwardly distorted form of closed loops at the lower right and upper left, this leads to centrifugal stretching and eventually breakout.

In both panels, the density shows a nearly spherical form in the DM within $R_{\rm K}$, but transitions to equatorial concentration just beyond $R_{\rm K}$. 
In the $x=0$ plane this equatorial material is 
trapped by closed loops, 
while in the $y=0$ plane it can flow along the field toward the loop top, where it stretches the field toward breakout.

The series of final five snapshots in figure \ref{fig:ZXZYlast50} illustrate the time sequence of mass build up, field line stressing, and centrifugal breakout.
The upshot is that material above $R_{\rm K}$ is confined in wings near the y-axis,
but escapes in perpendicular directions, leading to the lower-density gaps between these wings.

\subsection{Implications for Rigidly Rotating Magnetosphere model}

The plasma concentration into a warped-wing form, and the associated centrifugal distortion of the confining magnetic field, both have implications for the Rigidly
Rotating Magnetosphere (RRM) model introduced by \citet{TowOwo2005}.
% Townsend and Owocki (2005).
As with the  Rigid-Field Hydrodynamics (RFHD)
%2007MNRAS.382..139T 
formalism 
subsequently developed by \citet{Tow2007}, 
%Townsend et al. (2007),
this approach assumes the magnetic field is so dominant that it acts like rigid pipes that channel outflowing wind plasma to accumulation surfaces, set by minima in the combined centrifugal and gravitational potential.
This rigid field notion seems well justified by the very high wind-magnetic confinement parameters 
( $\eta_\ast > 10^6$) inferred for many B-stars with moderate to rapid rotation.

However, recent analysis \citep{2020MNRAS.499.5366O} for how centrifugal breakout likely sets the limit for CM
mass build up implies that, in practice, the confining fields are not in fact rigid, but rather are distorted by centrifugal forces near the mass accumulation limit set by breakout.

\begin{figure}
    \centering
\includegraphics[width=0.47\textwidth]{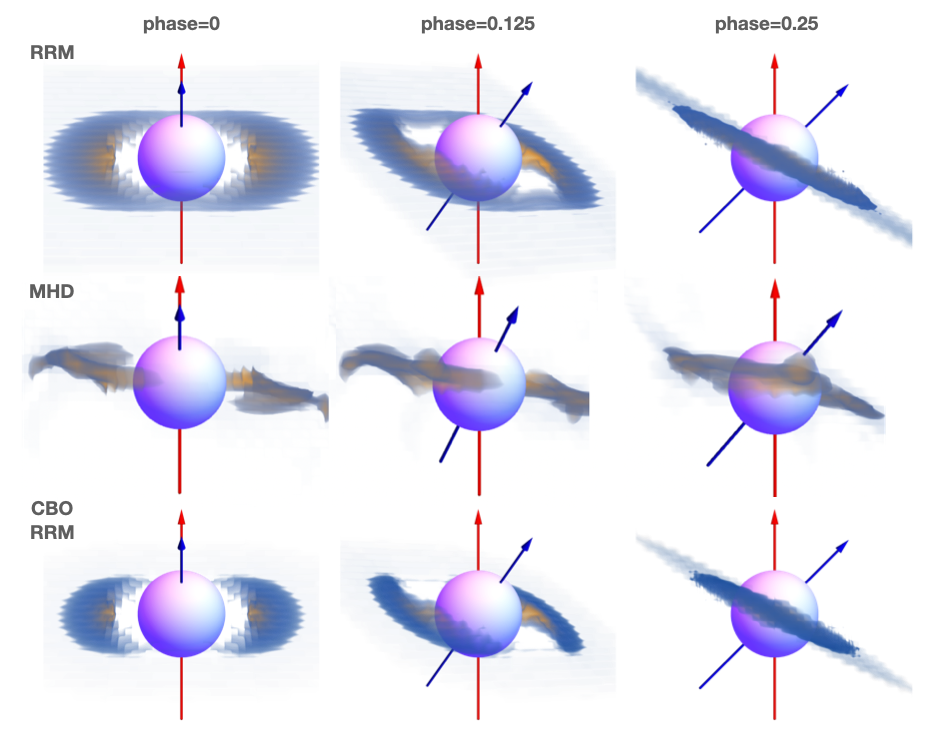}
\caption{
For standard oblique dipole case ($\beta =45^o$ and $W=0.5$), comparison of the phase variation of volume-rendered density for the final time ($t=1940$\,ks) of the MHD simulation (center row) with predictions of the analytic Rigidly Rotating Magnetosphere (RRM) model.
The top row depicts the density distribution resulting from wind feeding over a fixed time, as assumed in the original RRM analysis of \citet{TowOwo2005}.
The bottom row shows a modified scaling to mimic predictions of a centrifugal breakout (CBO) analysis, with higher concentration along the common magnetic-rotational equator (y-axis) and a radial decline in surface density, $\sigma \sim 1/r^5$, that is steeper than the $\sigma \sim 1/r^3$ for fixed-time wind-feeding along a dipole field.
The blue vectors depict the magnetic dipole axis, and the observer inclination is $i=90^o$, and so perpendicular to the red vectors representing the fixed stellar rotation axis.
}
\label{fig:4s}
\end{figure}

In this context, figure \ref{fig:4s} compares equatorial views of the projected density distribution of two different versions of the RRM model (top, bottom) with the results of the final time of the present MHD simulations (middle), for rotation phases 0, 0.125, and 0.025.
The RRM model in the top panel follows the original description from \citet{TowOwo2005}, in which the local density along the accumulation surface is just set proportional to local feeding rate by the stellar wind along that field line. 
Since the area of the flow tube scales inversely with the field strength, $A \sim 1/B$, mass flux conservation for a dipole field gives a surface density that declines with radius as $\sigma \sim B \sim 1/r^3$.
For a tilted dipole, the inner edge of the CM is closest to the star (roughly at $R_{\rm K}$) along the line (y-axis) of common magneto-rotation equator, giving it a somewhat higher density compared to other azimuths.
However, as shown in the top row of figure \ref{fig:4s}, the overall azimuthal variation in density is modest, much less than from the distinct wing structure of the MHD model in the middle row.

There are also differences in the 3D form of the RRM surface vs. the MHD wings,
with the former being distinctly offset from the rotational equator, and latter warped about that equator.
As a result, there are quite notable differences in the phase variations, for example in the timing and degree of occultation of the star by the respective CM's.

Moreover, much as found in the CBO analysis of the aligned rotation case \citep{2020MNRAS.499.5366O}, the radial decline in surface density in this MHD simulation of the oblique rotator is much steeper than the assumed $\sigma \sim B \sim 1/r^3$ scaling of original RRM analysis, 
instead following closer the CBO scaling with magnetic tension, $\sigma \sim B^2 \sim 1/r^6$.

To account for this, as well as the stronger azimuthal variation, the bottom panel of figure \ref{fig:4s} show an RRM
with density following a CBO-adjusted form given by eqn. (2) of  \citet[][]{2022MNRAS.511.4815B},
reproduced below as
 eqn.\, (\ref{eq:berry2}).
Specifically, this uses their standard value for radial power index $p=5$,
along with a scaling parameter $\chi = 0.1$, which sets the azimuthal decline in CM density away from the common magneto-rotational equator\footnote{The $\chi=0.1$ used  here is twice the value assumed by \citet{2022MNRAS.511.4815B}, giving a somewhat weaker drop in density away from the common equator. The original RRM model effectively assumes $\chi \gg 1$.}.
While the geometric differences remain, there is now an improved correspondence in the density distribution, and the associated occultations of the star.

\subsection{Distribution of Mass flux and Density}

\begin{figure}
\centering
\includegraphics[width=0.47\textwidth]{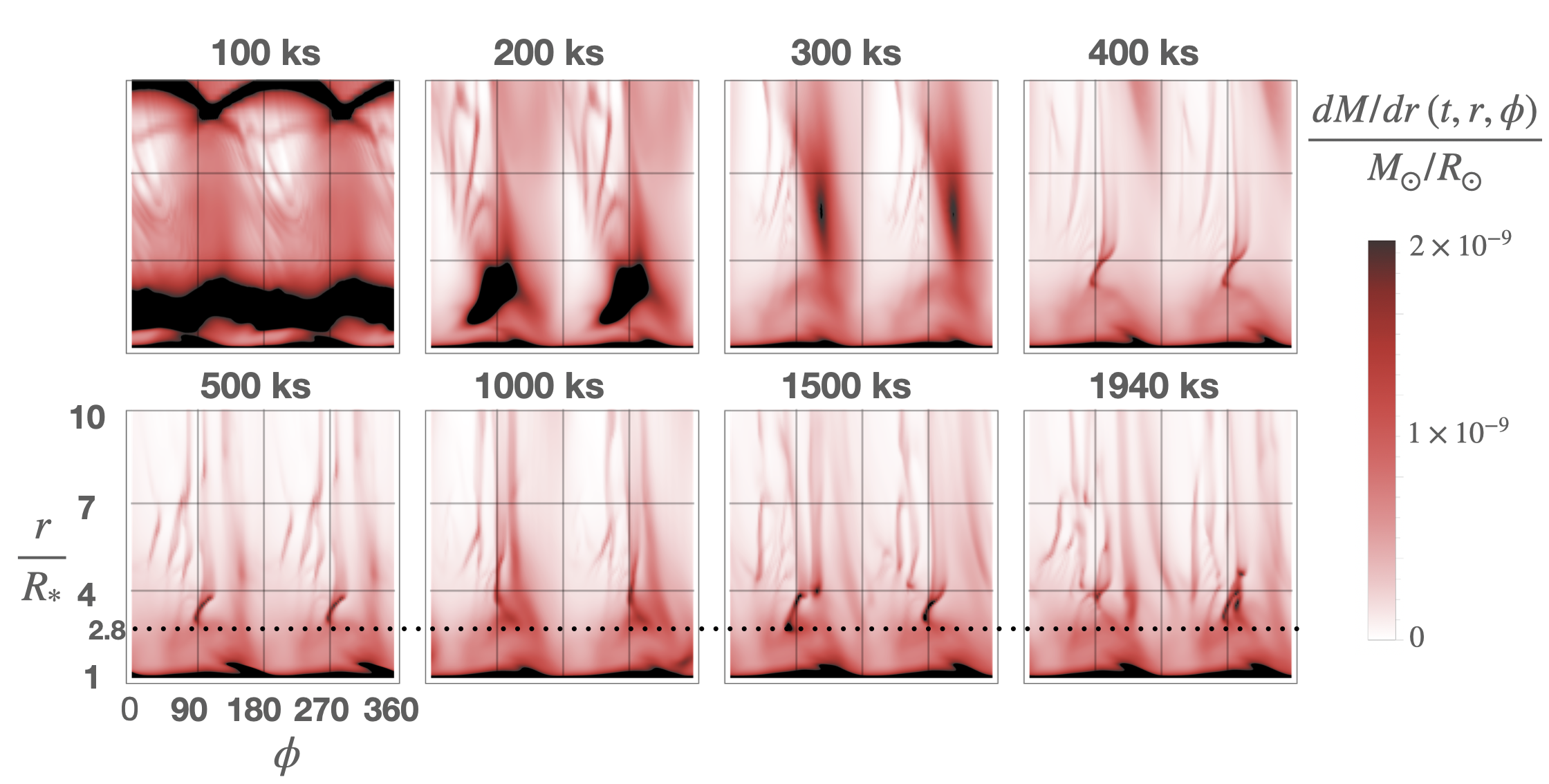}
   \caption{For MHD simulation of tilted dipole $\beta=45^o$, time evolution of the latitudinally averaged mass distribution in radius $dM/dr$,
   plotted versus radius and azimuth. After about 500 ks, the model reaches a quasi-steady state with episodic mass ejections distributed about  the y-axis ($\phi=90^o$ and $\phi=270^o$), where the magnetic and rotational equatorial planes intersect. 
   The horizontal dotted line at $r=2.8 R_\ast$ marks the boundary between magnetically confined material, and the onset of CBO events.}
    \label{fig:7}
\end{figure}

To give further insight into the overall structure and evolution of the MHD simulations for the tilted dipole ($\beta =45^o$) case, figure \ref{fig:7} 
shows the time evolution and spatial variation of the latitudinally integrated mass distribution in radius  $dM/dr$,
plotted versus radius $r$ and azimuth $\phi$ for the time snapshots denoted.
From initial development over times $100-400\,ks$ shown along the top row, the structure settles in a quasi-steady form, with episodes of mass ejection distributed about the y-axis positions ($\phi=90^o$ and $\phi=270^o$) representing the intersection between the magnetic and rotational equators.
The horizontal dotted line at $r=2.8 R_\ast$ marks the boundary between magnetically confined material, and the onset of CBO events.

\begin{figure}
    \centering
    \includegraphics[width=0.47\textwidth]{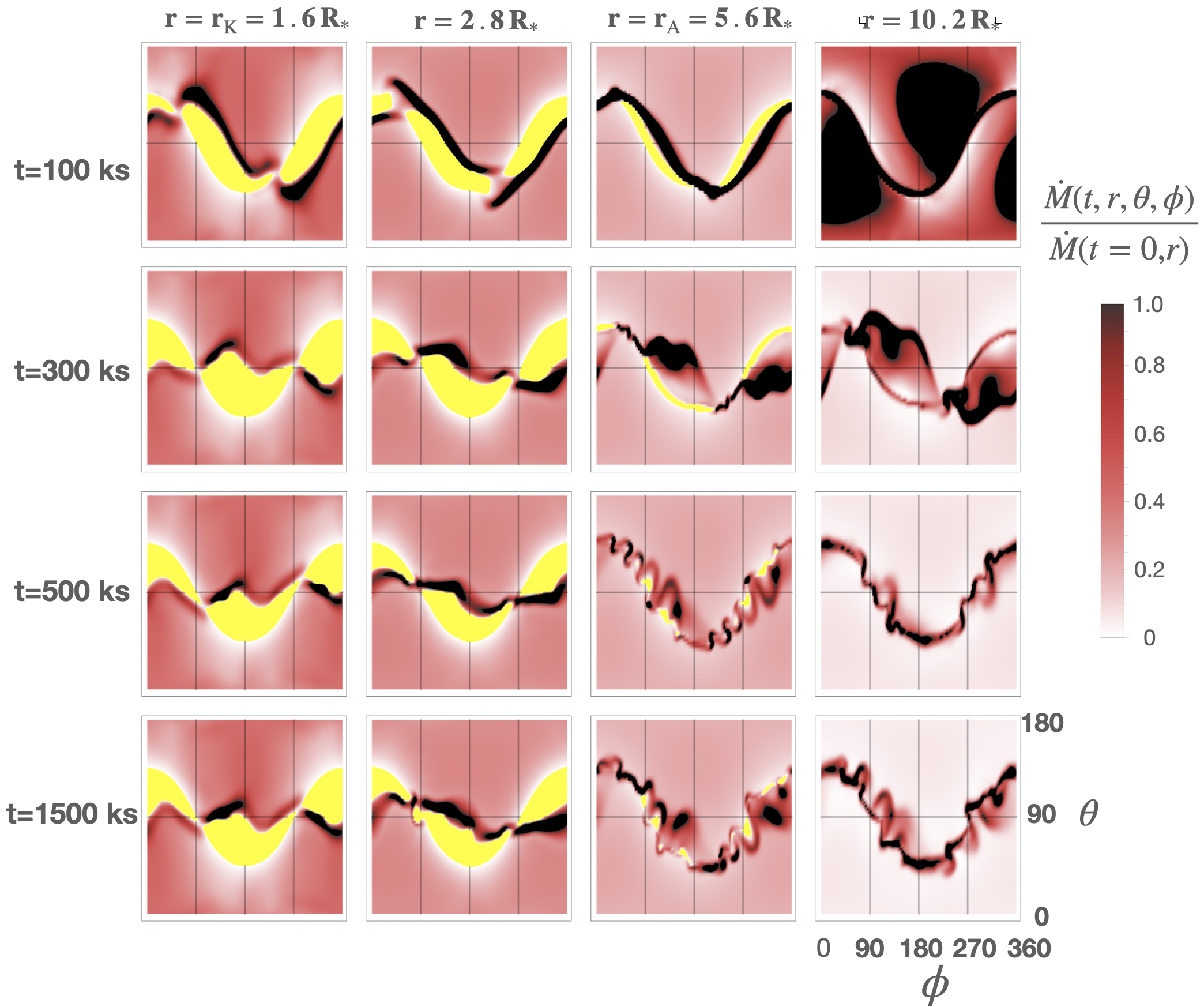}
   \caption{For tilted dipole $\beta=45^o$, mass flux plotted as a function of the azimuth and co-latitude at various radii. The columns show slices at the 4 labeled radii, while rows show snapshots at the 4 labeled times. Yellow denotes regions of infall.}
    \label{fig:8}
\end{figure}

Figure \ref{fig:8} shows mass flux plotted as a function of the azimuth and co-latitude at the labeled radii and times.
The yellow shows regions of infall that surrounded dark region of compressed outflow; at radii at and below the confinement radius $r=2.8 R_\ast$ this compressed outflow is near the {\em rotational} equator, but at larger radii, it shifts closer to the {\em magnetic} equator.

\begin{figure}
    \centering
\includegraphics[width=0.47\textwidth]{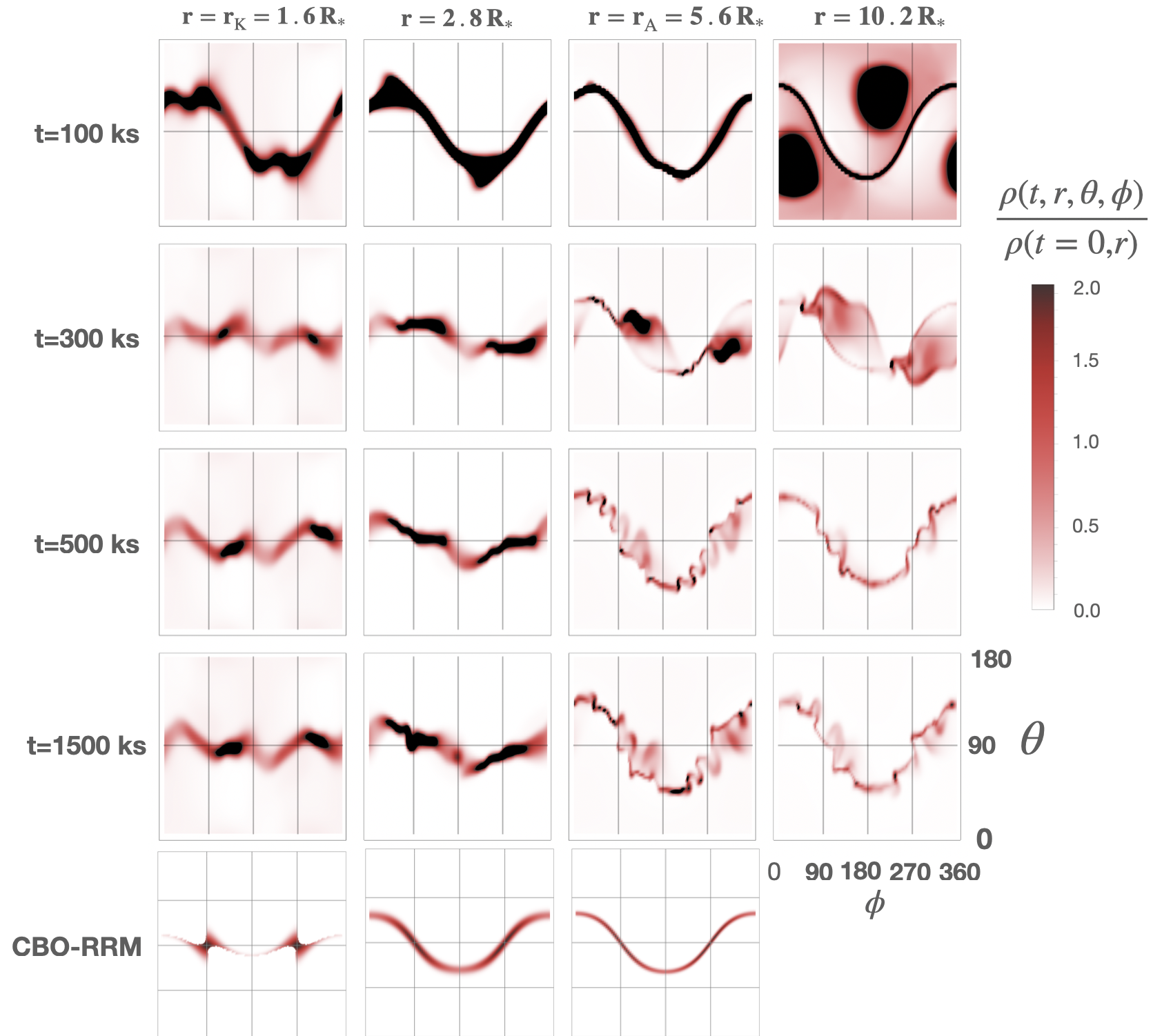}
   \caption{For tilted dipole $\beta=45^o$, radial slices of density plotted as a function of azimuth and co-latitude at the various radii labeled at the top, with time progression of MHD model in four rows at labeled time snapshots.  The bottom row shows for comparison the corresponding density distribution of the CBO-RRM model.}
    \label{fig:4}
\end{figure}

For the same samples in radius and time, Figure \ref{fig:4} now shows the  density, again plotted as a function of the azimuth and co-latitude.
This again shows that material is compressed near the rotational equator at radii at and below the confinement radius $r=2.8 R_\ast$, but closer to the magnetic equator at larger radii.
The bottom row compares the density for the CBO-RRM model, showing that the density concentration near $r=2.8 R_\ast$ is, in contrast to MHD result,   {\em intermediate} between the magnetic and rotational equators.

\section{Discussion and Future Work}

\subsection{Result summary
}

The central aim of this paper is to use 3D MHD simulations to characterize the centrifugal magnetospheres (CM) of strongly magnetic, rapidly rotating hot-stars for which the assumed dipole field has a significant tilt ($\beta = 45^o$) to the star's rotation axis.
Starting from an initial condition with a spherically symmetric, line-driven stellar wind, the MHD trapping and corotation of the wind outflow leads over many dynamical flow times to gradual build-up of material into a complex 3D CM, characterized by distinct {\em wings} of enhanced density, roughly centered on the line intersecting the
magnetic and rotational 
equatorial planes.
The asymptotic, quasi-steady-state includes repeated, small-scale centrifugal breakout (CBO) events, roughly centered about the direction of common equator, through which the ongoing wind feeding of the CM is balanced by CBO mass ejections.

The geometry of this dynamically fed CM follows roughly the minimum potential surfaces derived by the hydrostatic, rigidly rotating magnetosphere (RRM) model developed by \citet{TowOwo2005}, with however some key differences.
In particular, the surface density follows a steeper $\sigma \sim 1/r^5$ radial decline, reflecting the similar drop in magnetic tension $B^2 \sim 1/r^6$,
in contrast with the $\sigma \sim B \sim 1/r^3 $ scaling assumed for the original RRM model.
Moreover, the density is more concentrated azimuthally, into two wings centered on the common equatorial axis.
Both effects can be roughly captured by the parameterization introduced by \citet[][their eqn.\ 2]{2022MNRAS.511.4815B},
in which the surface density at a minimum potential location with radius $r$ and magnetic co-latitude $\theta_o$ is given by 
\beq
\sigma (r,\theta_o ) 
= \sigma_K 
\left ( \frac{R_K}{r} \right )^p \,
\exp (-\cos^2 \theta_o/\chi )
\, .
\label{eq:berry2}
\eeq
Here the surface density at the Kepler radius $R_K$ is given in terms of the magnetic field and gravity there,
\beq
\sigma_K = 0.3 
\frac{B_K^2}{4 \pi g_K}
\, .
\label{eq:berry6}
\eeq
Specifically, the comparisons in figure \ref{fig:4s} show that adopting $p=5$ and $\chi=0.1$ gives an overall density distribution (bottom row) that agrees better with MHD results (middle row) than the standard RRM result (top row).

But this figure also shows that the overall geometric form of the {\em dynamical} CM in the middle panel has some moderate deviations from the minimum-potential, {\em hydrostatic} accumulation surface assumed in even the CBO-modified RRM model shown in the lowermost panel.
This reflects the fact that, in contrast to the perfectly rigid dipole field assumed in the RRM paradigm, the dynamical CM naturally builds up to a limiting density that distorts this initial dipole, culminating in episodic CBO events and associated magnetic reconnection.

This field distortion leads to an associated {\em dynamical contortion} of the CM.
Instead of following the minimum total potential surface that generally lies {\em between} the magnetic and rotational equators, the inner regions of the dynamical CM lie closer to the {\em rotational} equator.
However, in the outer regions this transitions to a dense wind outflow that is concentrated toward the {\em magnetic} equator, and the associated wind current sheet that separates regions of opposite magnetic polarity.

\subsection{Open questions and future work}

Within these interesting new results and insights into the dynamical form of CM's, there remain several outstanding questions, grounded in limitations and approximations of these 3D MHD sims.

For example, the Courant limit on the time-step imposed by Alfv\'{e}n propagation across grid cells has so far limited the simulations to only moderately strong magnetic confinement parameter $\eta_\ast \lesssim 10^3$, much smaller than the $\eta_\ast \gtrsim 10^6$ estimated for known CM stars like $\sigma$~Ori~E.
In the associated stronger, stiffer magnetic field, it is possible that the dynamical  distortion effects identified here would be less pronounced. On the hand, in the view that this distortion stems from the inexorable build-up of CM density toward breakout, instead of the direct competition between field and wind outflow, then the CM contortion derived here may well be applicable to observed  CM stars. 
To distinguish between these different pictures, future work should carry out a parameter study in $\eta_\ast$, including extension to strong confinement, e.g., $\eta_\ast \lesssim 10^4$.

Future work should also explore a broader range of field tilt angles, including the extreme case of fully oblique dipoles, $\beta \rightarrow 90^o$,
which RRM analyses show to have a distinct ``cone-sheet'' form for the minimum-potential surfaces \citep{TowOwo2005}.
This sheet represents the asymptotic form of ``leaves" that form at large tilt angles, and it will be of interest to determine if these localized minima show plasma accumulation in full MHD simulations.

A further priority will be to derive observational diagnostics. 
For example, 
the RRM model predicts quite distinctive dynamical spectra for the rotational modulation of H$\alpha$ line emission \citep{Tow2005},
and it will be interesting how this may be altered by the dynamical distortion effects found in these MHD models.
It will also be of interest to see if CBO-induced magnetic reconnection events in the MHD models can reproduce the empirical scaling of incoherent, circularly polarized radio emission in massive stars \citep{2021MNRAS.507.1979L,2022MNRAS.513.1429S, 2022MNRAS.513.1449O} with potential implications for radio emission in Hot Jupiters \citep[e.g.][]{2017MNRAS.469.3505W}.

Synthesis of X-ray emission will require replacing the isothermal models here with a full energy equation. A key issue regards the outliers found  by \citet{Naze2014} in their correlation of observed X-ray luminosity with predictions from the dynamical magnetosphere (DM) model that applies for slow stellar rotation \citep{Owocki16}.  These outliers generally have relatively rapid rotation, and so are better modeled as having CM's than DM's.  A key question is whether the stronger observed X-rays might arise from stronger shocks with a higher duty cycle in CM's than DM's, or whether the CBO-induced magnetic reconnection might contribute to the inferred enhanced X-rays.

Finally, in our focus here on the dynamical form of the CM in these 3D MHD simulations, we have not yet examined the loss of angular momentum associated with the magnetic stresses and mass outflow in open field regions. For the field-aligned case ($\beta=0$), MHD models have provided a simple analytic scaling law for how this angular momentum loss scales with magnetic field strength, mass loss rate, and stellar rotation 
\citep{udD2009}.
But a key, open question, so far  only tentatively explored for 3D MHD models with modest magnetic confinement parameter
\citep{2022MNRAS.515..237S},
is how the non-zero tilt angle between the magnetic and rotation axes
might alter this spindown scaling law. This will thus be a central focus of planned parameter studies of models with a range of tilt angles $\beta$.

\section*{Data Availability Statement}
The data underlying this article will be shared on reasonable request to the corresponding author.
\section*{Acknowledgements}

This work is supported in part by the National Aeronautics and Space Administration under Grant No. 80NSSC22K0628 issued through the Astrophysics Theory Program.
AuD and MRG acknowledge support by the National Aeronautics and Space Administration through Chandra Award Numbers TM-22001 and GO2-23003X, issued by the Chandra X-ray Center, which is operated by the Smithsonian Astrophysical Observatory for and on behalf of the National Aeronautics Space Administration under contract NAS8-03060. This work used the Bridges2 cluster at the Pittsburgh Supercomputer Center through allocation
AST200002 from the Extreme Science and Engineering Discovery Environment (XSEDE), which was supported by National Science Foundation grant number 1548562.

\bibliographystyle{mnras}
\bibliography{udDoula}{}
%\begin{thebibliography}{}
%\makeatletter

%\makeatother
%\end{thebibliography}
\end{document}